\documentstyle[12pt,epsfig]{article}

\newcommand{\beq}{\begin{equation}}
\newcommand{\eeq}{\end{equation}}
\newcommand{\nn}{\nonumber}
\newcommand{\bea}{\begin{eqnarray}}
\newcommand{\eea}{\end{eqnarray}}

\newcommand{\gsim}{\ \rlap{\raise 2pt\hbox{$>$}}{\lower 2pt \hbox{$\sim$}}\ }
\newcommand{\lsim}{\ \rlap{\raise 2pt\hbox{$<$}}{\lower 2pt \hbox{$\sim$}}\ }

\newcommand{\np}[1]{Nucl. Phys. {\bf #1}}
\newcommand{\pl}[1]{Phys. Lett. {\bf #1}}
\newcommand{\pr}[1]{Phys. Rev. {\bf #1}}
\newcommand{\prl}[1]{Phys. Rev. Lett. {\bf #1}}
\newcommand{\zp}[1]{Z. Phys. {\bf #1}}
\newcommand{\prep}[1]{Phys. Rep. {\bf #1}}

\newcommand{\mpl}[1]{Mod. Phys. Lett. {\bf #1}} 
\newcommand{\ptp}[1]{Prog. Theor. Phys. {\bf #1}}

\makeatletter
\if@twoside
\else \oddsidemargin 00pt \evensidemargin 00pt
\fi
\let\@eqnsel = \hfil

\expandafter\ifx\csname mathrm\endcsname\relax\def\mathrm#1{{\rm #1}}\fi
\@ifundefined{mathrm}{\def\mathrm#1{{\rm #1}}}{\relax}

\makeatother

\begin{document}
\thispagestyle{empty}
\null
\hfill FTUV/96-42, IFIC/96-50

\hfill hep-ph/9607281

\vskip 1.5cm

\begin{center}
{\Large \bf      

Non-decoupling and lepton number violation in left-right models

\par} \vskip 2.em
{\large		
{\sc G. Barenboim  and M. Raidal
}  \\[1ex] 
{\it  Departament de F\'\i sica Te\`orica, Universitat 
de Val\`encia}\\ 
{\it and IFIC, Centre Mixte Universitat 
de Val\`encia - CSIC} \\
{\it E-46100 Burjassot, Valencia, Spain} \\[1ex]
\vskip 0.5em
\par} 
\end{center} \par
\vfil
{\bf Abstract} \par
We argue that large non-decoupling effects 
of heavy neutrinos can appear naturally in 
manifestly left-right symmetric models due to the 
minimization conditions of the scalar potential and the 
structure of vev's imposed by  phenomenology. 
We derive constraints on off-diagonal light-heavy and heavy-heavy
neutrino mixings from the searches for lepton violating decays
$\mu\rightarrow e\gamma,$ $\mu\rightarrow e e^-e^+$ and 
$\mu-e$ conversion in nuclei. The most stringent limits come from the 
latter process because its amplitude shows a quadratic
non-decoupling dependence on the heavy neutrino mass. 
Due to the suppression of right-handed currents by large $W_R$
mass the present experiments are not sensitive to the 
intergenerational mixings between heavy neutrinos 
if $M_{W_R}>200$ TeV.

\par
\vskip 0.5cm
\noindent July 1996 \par
\null
\setcounter{page}{0}
\clearpage

\section{Introduction}

Although  the current experimental data are  consistent with the
standard model of weak  and electromagnetic interactions (SM)
with  an impressive accuracy  there are several hints
that  the SM is  not  the ultimate
theory of Nature. Indeed, the anomalies measured in the solar
\cite{sol} and atmospheric \cite{atm} neutrino fluxes seems
to require that neutrinos have a tiny mass, manifested in these phenomena
through flavour oscillations. The observations of COBE satellite \cite{cobe}
indicate the existence of a hot neutrino component of dark matter.
These phenomena  cannot be explained in the framework of the SM
and, therefore,  motivate us to search for   physics
 beyond the electroweak scale. 

 The new physics at higher energies should incorporate
new heavy states   
which, in general, mix with the light ones. These  mixings can 
occur, for example, in the neutral fermion sector
 between light and heavy neutrinos  or in the gauge boson sector
between the SM gauge bosons and the gauge bosons of  
the new  interactions
modifying slightly the SM couplings of 
 charged and neutral currents.  
Such effects have been searched for using collider as well as  low energy
precision data. Since no deviations from the SM predictions have been found
one has been able  to constrain the 
mixing angles  and masses of the new particles considerably
 \cite{lang,precision,lang2}.    
In the following discussion
we shall concentrate on the properties of heavy 
neutrinos i.e.  neutral, weakly interacting
fermions with masses larger than  $M_Z.$

Motivated by the effective field theory approach 
it is natural to expect that   heavy particles  decouple from low
energy life, i.e.  the effects of heavy 
particles in the virtual intermediate states would be
suppresed by inverse powers of the heavy scale.
Indeed, since the low energy theory of any extended gauge model 
(where the new heavy states are singlets of the SM gauge group)
should be  the SM then due to its renormalizability the 
heavy scale decouples \cite{senj2}.
The above  result was first derived by Appelquist and Carazzone \cite{nondec}. 
However, large effects of the heavy states much below their
production threshold are possible. They occur due to the loop corrections 
which grow with the   mass of  heavy particle to some power. 
These non-decoupling effects have been studied previously in many
works \cite{loop} 
in which  loops involving heavy neutrinos have also been considered.
If the smallness of ordinary neutrino mass
is explained by the see-saw mechanism \cite{see-saw} then
the non-decoupling effects 
are cancelled by the small see-saw mixing angles. 
However, if the see-saw mechanism is not active then 
these  effects can be 
significant. 

Non-decoupling effects are, indeed, phenomenologically very interesting 
because
they allow to explore the physics at  high  scales through low
energy processes.
To enhance the effects we shall consider  possibilities 
other than the see-saw mechanism to
keep the masses of the known neutrinos below the laboratory limits
and at the same time 
allow   mixing angles between  heavy and light 
neutrino states to be large.
Such alternative scenarios, where vanishingly small neutrino
masses are ensured by some symmetry arguments  have been considered  
in many works \cite{wyler,esingl,buch}. 
Some of them,  motivated by the generalized $\rm E_6$ models, suggest
a very specific form of neutrino  mass matrix involving  
a large number of new heavy doublet and singlet neutrinos \cite{esingl}.
 The others demand relations
for  neutrino mass matrix  \cite{buch} which cannot be
reasonably justified in the SM enlarged 
by adding  right-handed singlet neutrinos. 
Since the only energy scale involved  in the SM
 is the electroweak breaking scale the latter
case, thus, introduces a fine tuning for Yukawa couplings in the neutrino
sector.  

However, this is not the case in models with extended gauge sector.
One of the most interesting extensions of the SM is based on
the gauge group $SU(2)_L\times SU(2)_R\times U(1)_{B-L}$ \cite{lr}.
At the Lagrangian level this model is left-right symmetric but,
in order to explain the observed 
parity violation of the weak interaction,
the  vacuum is not invariant under the left-right symmetry. 
The left-right  model with triplet 
representations of Higgs fields can accomodate
the see-saw mechanism for neutrino masses \cite{ss}. 
As we are going to argue,
due to the built in hierarchy of  vacuum expectation values (vevs) of the 
Higgs fields and
 relations between the vevs obtained in minimization of the Higgs potential,
large non-decoupling effects of heavy neutrinos 
can occur in the model with manifest left-right symmetry.

In this work we are going to constrain the off-diagonal elements
of  light-heavy and  heavy-heavy neutrino mixing matrices from the 
lepton flavour violating processes $\mu\rightarrow e\gamma,$
$\mu\rightarrow e e^- e^+$ and
$\mu - e$ conversion in nuclei   in the manifestly 
left-right symmetric model. 
These processes involve virtual  heavy, predominantly right-handed,
neutrinos and gauge bosons as well as their mixing angles with the 
light states.
The amplitudes of the two latter processes grow with heavy neutrino mass
to the second power  showing a genuine non-decoupling dependence.
This non-decoupling behaviour is comparable to the top mass dependence
of the $\rho$ parameter and of the $Z \rightarrow b \overline{b}$ 
vertex.
 We show that 
the perturbative unitarity bound on
heavy neutrino mass in the left-right model is not as restrictive as it is in 
the $\rm E_6$ based extensions of the SM \cite{tommde}
and,  therefore, experiments can bound the flavour changing mixings   more
strictly than  in the latter case.

The outline of the paper is the following. In  Section 2 we present 
basics of the left-right symmetric model and discuss the mechanism
for natural  occurrence of large mixings of heavy neutrinos. 
In Section 3 we consider the dominant decays of heavy neutrinos and 
derive the perturbative unitarity 
condition for   neutrino masses.
In Section 4 we calculate the branching ratios of the
 processes $\mu\rightarrow e\gamma,$  $\mu\rightarrow e e^- e^+$
and $\mu - e$ conversion in nuclei. We discuss the non-decoupling effects 
and derive limits on the off-diadonal light-heavy and heavy-heavy 
neutrino mixings. 
Our conclusions are given in Section 5.

\section{Light-heavy mixings in left-right symmetric models}

We begin with presenting  the minimal $SU(2)_L \times SU(2)_R 
\times U(1)_{B-L} $ model with a left-right discrete symmetry. 
In the left-right symmetric models each generation of quarks and
leptons is asigned to the multiplets
\bea
Q = \pmatrix { u \cr d \cr} \; \; , \; \; \psi=\pmatrix{ \nu \cr l \cr} \;\; ,
\eea
with the quantum numbers $(T_L,T_R,B-L)$
\bea
Q_L \; :\; \left( \frac{1}{2},0,\frac{1}{3} \right) \;\;\;  , \;\;\; 
\psi_L \; :\; \left( \frac{1}{2},0,-1 \right) , \nn \\
Q_R \; :\; \left( 0,\frac{1}{2},\frac{1}{3} \right)  \;\;\;  , \;\;\; 
\psi_R \; :\; \left(0,\frac{1}{2},-1 \right) .
\eea

Concerning the Higgs sector, in order to give masses 
to fermions, all the left-right models should 
contain a bidoublet
\bea
\phi = \pmatrix {\phi_1^0 & \phi_1^+ \cr \phi_2^- & \phi_2^0 \cr}=
(\frac{1}{2}, \frac{1}{2}^\ast, 0).
\eea
Since the bidoublet does not break the right-handed symmetry
the Higgs sector has to 
be enlarged. This procedure is not unique but interesting models are 
obtained by adding the  scalar triplets
\bea
\Delta_{L,R} = \pmatrix {\frac{\Delta_{L,R}^+}{\sqrt{2}}  & 
\Delta_{_{L,R}}^{^{++}} \cr 
\Delta_{_{L,R}}^{^0} & \frac{-\Delta_{L,R}^+}{\sqrt{2}}\cr}
\eea
with the quantum numbers $ 
\Delta_L \;: \; (1,0,2) $ and $ \Delta_R \; : \; (0,1,2) , $
respectively.
In addition, we require the full Lagrangian of the model to be 
manifestly left-right symmetric i.e. invariant under 
the discrete symmetry
\beq
\psi_L \longleftrightarrow \psi_R \;,\;\; 
\Delta_L \longleftrightarrow \Delta_R  \;,\;\;
\phi \longleftrightarrow \phi^\dagger. 
\label{trans}
\eeq
This symmetry plays a role in minimizing the Higgs potential
and, therefore, 
is crucial for obtaining the non-decoupling effects in the model.

In general, our symmetry breaking would be triggered by the  vevs
\bea
\langle \phi \rangle  = 
\pmatrix {\frac{k_1}{\sqrt{2}} & 0 \cr 0 & \frac{k_2}{\sqrt{2}}\cr}
\; \; \; \; , \; \; \; \;
\langle \Delta_{L,R} \rangle = \pmatrix {0   & 0 \cr 
\frac{v_{L,R}}{\sqrt{2}} & 0 \cr}.
\eea
The vev $v_R$ of the right triplet breaks the $SU(2)_R 
\times U(1)_{B-L} $ symmetry  to $U(1)_Y$ and gives masses to new
right-handed particles. 
Since the right-handed currents are not observed, $v_R$ should be
sufficiently large \cite{lang2,wmass}. 
Further, the vevs $k_1$ and $k_2$ of the
 bidoublet break the SM symmetry and, therefore, are of the order
of electroweak scale. The vev $v_L$ of the left triplet, which 
contributes to the $\rho$ parameter, is quite tightly bounded 
by experiments \cite{gun}
and should be below a few GeV-s.
Thus,  the following hierarchy should be satisfied:
 $ |v_R| \gg |k_1|, |k_2|  \gg |v_L| $.
In principle, due to the underlying symmetry 
two of the vevs can be chosen to be real
but two of them can be complex leading to the 
spontaneous $CP$ violation \cite{gabriela}. To simplify the 
discussion we assume in the following that all the vevs are real.

The most general Yukawa Lagrangian for leptons invariant under the gauge group 
is given by
\bea
-{\cal L}_Y &=& f_{ij} \bar{\psi_L^i} \phi \psi_R^j + 
	g_{ij} \bar{\psi_L^i} \tilde{\phi} \psi_R^j  + \mbox{h.c.} \nn \\
& & + i (h_M)_{ij} \left( \psi_L^{i \, T} C \tau_2 \Delta_L \psi_L^j  + 
 \psi_R^{i \, T} C \tau_2 \Delta_R \psi_R^j \right) + \mbox{h.c.} ,
\label{lag}
\eea
where $f$, $g$ and $h_M$ are matrices of Yukawa couplings.
The left-right symmetry (\ref{trans}) requires $f$ and $g$ to be
Hermitian. The Majorana couplings $h_M$ can be taken to be real and 
positive due to our ability to rotate $\psi_L$ and $\psi_R$ by a 
common phase without affecting $f$ and $g.$ 

According to the Lagrangian (\ref{lag}),
neutrino masses derive both from the $f$ and $g$ terms, which lead to 
Dirac mass terms, 
and from the $h_M$ term, which leads to large  Majorana mass terms.
Defining, as usual, $\psi^c \equiv C ( \bar{\psi})^T $, 
the mass Lagrangian following from Eq.(\ref{lag})  can be written in
the form
\bea
-{\cal L}_{mass}=\frac{1}{2}(\bar{\nu}^c_L\,M\,\nu_R +
\bar{\nu}_R\,M^\ast\,\nu^c_L)\,,
\eea
where $\nu_L^c=(\nu_L, \nu_R^c)^T$ and $\nu_R=(\nu^c_L, \nu_R)^T$ are
six dimensional vectors of neutrino fields.
The neutrino mass matrix $M$ is complex-symmetric and
can be written in the block form
\bea
M =  \pmatrix{ M_L & M_D \cr 
	M_D^T & M_R \cr},
\label{matr}
\eea
where the entries are $3\times3$ matrices given by 
\bea
M_L=\sqrt{2}h_M v_L\,,\;\; M_D=h_D k_+\,, \;\;
M_R=\sqrt{2}h_M v_R\,.
\label{sub}
\eea
Here we have defined $h_D=(f k_1 + g k_2)/(\sqrt{2} k_+),$ where 
$k_+^2=k_1^2+k_2^2.$
 The masses of charged leptons are
given by $M_l=(g k_1 + f k_2)/\sqrt{2}$ and, therefore,
without fine tuning of $f$ and $g$ one has $ M_D\simeq M_l.$
Moreover, on the basis of avoiding possible fine tunings 
it is natural to assume that all the Yukawa couplings $h_{M,D}$
are of similar magnitude for a certain lepton family.
In this case the mass matrix (\ref{matr}) has a strong
hierarchy between different blocks which is set by the 
hierarchy of vevs. 

Since the neutrino mass matrix is symmetric it can be diagonalized 
by the complex orthogonal transformation
\bea
U^T\, M \, U=M^d,
\eea
where $M^d$ is the diagonal neutrino mass matrix. If we denote 
\bea
U=\pmatrix{U^\ast_L \cr U_R \cr},
\eea    
then in the basis where the charged lepton mass matrix is diagonal
(we can choose this basis without loss of generality) 
the physical neutrino mixing matrices  which appear
in the left- and right-handed charged currents are simply  given by
$U_L$ and $U_R,$ respectively.
Indeed, $U_L,$ $U_R$ relate the left-handed and right-handed neutrino 
flavour eigenstates $\nu_{L,R}$ with  the mass eigenstates $\nu_m$ 
according to
\bea
\nu_{L,R}=U_{L,R}\,\nu_m\,.
\label{invmatr}
\eea

In order to find relations between 
the different parameters of the theory
we have to solve the first derivative equations of the
scalar field potential $V$
\bea
\frac{\partial V}{\partial v_{L,R}}=0\,,\;\;
\frac{\partial V}{\partial k_{1,2}}=0\,.
\eea 
The full potential, invariant under the transformation (\ref{trans}),
can be found in Ref.\cite{gun,gabriela}.
When considering the minimization of the potential 
we get a see-saw type relation among the vevs 
\bea
v_L v_R \simeq k_1 k_2\,. 
\label{potrel}
\eea
Here we have excluded a possibility of fine tuning and  
assumed that all the parameters of the potential are of the same
order.

Having this in mind we can now start to study  mass matrices 
of type (\ref{matr}). For a simple one family case we find 
\bea
\mbox{det} M\sim 2 v_L v_R - k_+^2 = 2 v_L v_R - (k_1^2 +k_2^2)\,.
\eea
If $k_1 \neq k_2 $, then Eq.(\ref{potrel}) implies that
$M$ is not singular and we have the usual see-saw mechanism. 
It is convenient to
employ the self-conjugated spinors
\bea 
\nu \equiv \frac{\nu_L + \nu_L^c}{\sqrt{2}}  \, , \; \; \; 
N \equiv \frac{\nu_R + \nu_R^c}{\sqrt{2}}\,,
\eea
which are also the approximate mass eigenstates with masses
\bea
m_{\nu} & \simeq & \sqrt{2} \left( h_M v_L - \frac{h_D^2 k_+^2}{2 h_M v_R}
\right)\,,
\nn \\
m_N & \simeq & \sqrt{2} h_M v_R\,, 
\eea
 respectively. The mixing between them
depends on the ratio of the masses  as 
$\sin^2\theta\sim m_\nu/m_N$  being vanishingly small for the present
experimental values of $m_\nu$ and $m_N$ \cite{pdb}. 

On the other hand, if $k_1 = k_2$, then it follows from Eq.(\ref{potrel})
that $M$ is singular.
Indeed, this is the most natural case since $k_1$ and $k_2$ are
 vevs of the same bidoublet. 
The discrete left-right symmetry (\ref{trans})  togheter with 
the imposition of the
discrete symmetry $\phi_1  \longleftrightarrow \phi_2$ 
\cite{senj} ensures
that the scalar fields $\phi_1$ and $\phi_2$ are indistinguishable 
at the Lagrangian level.
As they both acquire vevs at the same stage
of the symmetry breaking 
 it is natural to expect that also the vacuum
respects this symmetry.   
In this case the mixing angle
$\sin \theta \sim  k/v_R $ is no longer  related to the ratio
of the light and heavy neutrino masses 
and would be allowed to be as large as
$O(10^{-1}).$ The masses then, are given by
\bea
m_{\nu}  \simeq  0\,, & \;\;
m_N  \simeq   \sqrt{2} h_M v_R\,, 
\eea
which explains the smallness of the mass of the known neutrino.
According to this one generation example the singularity of the mass matrix,
indeed, may arise naturally in our model. However, it should be noted that 
it is still an imposed relation which can be justified by some higher 
symmetry.

This given example can be generalized
to the three generation model. The sub-matrices $M_{L,D,R} $
of the mass matrix (\ref{matr}) are proportional to the corresponding
vevs and, together with Eq.(\ref{potrel}), there is a natural structure
for a singular mass matrix with three times degenerate
zero eigenvalue. The remaining three neutrinos have masses of the 
order of $v_R.$ The neutrino
mixing matrixes $U_L$ and $U_R$ consist of  two $3\times3$
blocks. In $U_L$ one of them presents light-light and another
light-heavy mixings while $U_R$ contains light-heavy and heavy-heavy
mixings. According to our model, the light-light and heavy-heavy
mixing angles are naturally maximal while the light-heavy mixing angles are 
of the order of $ k/v_R.$
 This implies that the light-heavy mixings can also be
substantial. The intergenerational mixings among light neutrinos
can be constrained, for example, in neutrino oscillation experiments
\cite{sol,atm}.
In the present paper we are going to constrain the off-diagonal
light-heavy and heavy-heavy mixings from flavour 
 changing processes. 

\section{Perturbative unitarity bound on massive neutrinos}

As mentioned in  
the Introduction, due to the non-decoupling effects
in our model, the branching ratios of the flavour changing processes we are
considering are increasing functions of 
neutrino masses. Therefore, in order to
bound neutrino mixing angles from the existing measurements we must first
study the viable range of masses the heavy neutrinos can take in our model. 
In general, if neutrino is extremely massive its decay width becomes so
broad that one cannot identify it as a particle. In this way,
assuming that 
\bea
\Gamma \lsim \frac{m_N}{2}\,,
\label{per}
\eea
 where $\Gamma$ is the total width of 
heavy neutrino, one  can obtain a perturbative unitarity bound on 
the heavy neutrino mass. It turned out that in the SM with
right-handed singlet neutrinos this bound is very restrictive
demanding the heavy neutrino mass to be below ${\cal O}(1)$ TeV.
 This  is
an expected result since the unique energy scale in this model is the
electroweak breaking scale and, therefore, neutrinos cannot be  
considerably heavier. Consequently, the experimental 
bounds on off-diagonal heavy-light 
neutrino mixings in models of this type are 
affected by this result \cite{tommde}.

In left-right models there are  new right-handed 
interactions and a new energy scale associated with the 
breaking of the right-handed symmetry. Due to these new ingredients
the perturbative unitarity bound on neutrino mass is not
related any more to the scale of left-handed interactions but to the 
right-handed one. 
Therefore, in the left-right models one can expect to obtain more stringent
bounds on the light-heavy neutrino mixings than in the 
models with right-handed singlets.

The dominant decay modes of the heavy neutrino are 
\bea
N &\rightarrow & e+W_{R}\,, \label{d1} \\
N &\rightarrow & e+W_{L}\,, \label{d2} \\
N &\rightarrow & \nu+Z_L\,. \label{d3} 
\eea
We assume that possible decays to other heavy neutrinos and $Z_R$
or Higgs bosons are suppressed by phase space factors due to the
large masses of the final state particles. 
If $N$ is much heavier than $W_R$ then the first decay occurs
without any suppression while the latter two are supressed either 
by the gauge boson mixing angle (decay (\ref{d2}))
 or by light-heavy neutrino mixing angle (decay (\ref{d3})). Since
we are interested in obtaining an 
upper bound on the neutrino mass  then we assume that the
condition $m_N>M_{W_R}$ holds and we expect the decay (\ref{d1}) to be
dominant. 
The decay rates of  (\ref{d1}) and (\ref{d2}) can be written as
\bea
\Gamma_{W_{L,R}}=\frac{g^2}{64\pi M^2_{W_{L,R}}}\beta^2_{L,R}m^3_N
\left(1+\frac{M^2_{W_{L,R}}}{2m_N^2}\right)\left(1-\frac{M^2_{W_{L,R}}}{m^2_N}
\right)^2\,,
\label{der1}
\eea
where $\beta_R=1$ and $\beta_L=\xi$ is the gauge bosons
mixing angle. Similarly we obtain for the decay (\ref{d3})
\bea
\Gamma_{Z_L}=\frac{g^2}{64\pi M^2_{W_L}}\theta^2 m^3_N
\left(1+\frac{M^2_{Z_{L}}}{2m_N^2}\right)\left(1-\frac{M^2_{Z_{L}}}{m^2_N}
\right)^2\,,
\label{der2}
\eea
where $\theta$ is the mixing angle between $N$ and $\nu.$

Let us first analyse the decay (\ref{d2}). 
In the limit $M^2_{W_L}\ll m^2_N$ 
substituting the 
numerical values to Eq.(\ref{der1}) we get $m_N<30 \, M_{W_L}/\xi.$
In the left-right model, assuming $k_1=k_2,$ the mixing angle 
between  left and right gauge bosons is given by 
$\xi= M_{W_L}^2/M_{W_R}^2$ and we obtain
\bea
m_N<30\, \frac{M_{W_R}}{M_{W_L}} M_{W_R}\,.
\label{mn1}
\eea 
On the other hand, from the decay (\ref{d1}), which does not depend on
any mixing angle, we get in the limit $M_{W_R}^2\ll m_N^2$  
\bea
m_N<30 \, M_{W_R}\,,
\label{mn2}
\eea 
a much stronger requirement than the previous one. Indeed, since $\xi$ depends
on the gauge boson masses to the second power one  expects to
have  weaker bound from the decay involving $W_L$ than from that involving
$W_R.$ The right-hand side of Eq.(\ref{mn1}) is enhanced by a factor of
$M_{W_R}/M_{W_L}$ if compared with Eq.(\ref{mn2}) which means that,
 in the case of very heavy neutrino, the 
decay mode (\ref{d1}) dominates over (\ref{d2}). 

However, in models with non-decoupling of heavy neutrinos the
neutrino mixing angle $\theta$ is given by 
$\theta= k/v_R=M_{W_L}/M_{W_R},$ the ratio of the masses to the first power.
Taking into account the expression for neutrino mixing angle  we
 obtain from Eq.(\ref{der2}) exactly the  bound (\ref{mn2}) also 
for the decay mode (\ref{d3}). 
We know from precision measurements that the mixing of
light neutrinos with heavy neutral fermions 
are limited to $\theta^2_e < 0.005,$ 
$\theta^2_{\mu}< 0.002$ and $\theta^2_{\tau} < 0.01$  \cite{precision}, where  
the subscripts denote corresponding generations.
These limits are of the same order of magnitude as the limit
predicted by our model taking into account the present
lower bound on right-handed gauge boson mass $M_{W_R}\gsim 1.4$ TeV
\cite{wmass}.  
Consequently, if  non-decoupling is the true scenario
then for very heavy neutrinos the decay rates of the processes
(\ref{d1}) and (\ref{d3}) are almost equal in magnitude.
For lighter neutrinos (but still heavier than $W_R$) the
former decay can be suppressed by a phase space factor which implies
that the heavy neutrino decays (\ref{d3}) dominate.
However, if non-decoupling does not  occur then the decay mode (\ref{d1})
dominates. In any case, the perturbative unitarity bound on
heavy neutrino mass is with a good precision given by Eq.(\ref{mn2}). 
With this result we proceed to the analysis of lepton
flavour violating processes.

\section{Constrains on neutrino mixings from flavour changing processes}

Indirect limits on intergenerational light-heavy neutrino mixing
angles can be derived from the precision measurement limits on the diagonal 
mixing angles by using the Schwartz inequality. 
The resulting bounds are all in the
range of $(U_LU_L^\dagger)_{ab}< (0.003-0.004)$ \cite{tommde}, 
where $a,b$ stand for any generation index. They are
 more stringent than any direct bound from tree level processes
\cite{lalo}. 
Nevertheless, loop diagrams involving  heavy neutrinos 
give rise to unobserved rare processes such as $\l_i \longrightarrow
 \gamma l_j$, $l_i \longrightarrow l_j \bar{l}_j l_k $, etc. where $i$
is either a tau or a muon while $j$ and $k$ are electron or muon in the 
former case and only electron in the latter case.
Taking into account the indirect limits the rates for all the
processes involving  violation of the tau lepton number
turn out to be below the experimental sensitivity even for
extreme values of the heavy neutrino masses \cite{t22}.
However, due to the extraordinary sensitivity of the
experiments  looking for flavour changing processes involving the
first two families the obtained constraints are
significantly stronger than the indirect 
limits. Because of this we are going to consider the first
two families only.

Let us first consider the decay $\mu\rightarrow e\gamma$ induced 
in left-right models by 
the one-loop Feynman diagrams 
depicted in Fig.1.  
The additional contribution arising from the diagrams involving lepton
flavour violating triplet  Higgs bosons $\Delta^{++}$  have been 
studied in Ref.\cite{moha}. 
The triplet Higgs bosons must be heavy because their masses are set
by $v_R.$  Since there is no non-decoupling effect working
in these diagrams (lepton flavour violation is induced in the vertices
not in the internal lines) 
their contribution to the $\mu$ decay 
is supressed by $v_R^{-4}$ and can be neglected 
in our approach.  
The  dominant diagrams involve both left- and right-handed
charged current interactions and, consequently, both left-
and right-handed neutrino mixing matrices $U_{L,R}.$ 
The branching ratio for the process  takes the
form
\bea
B(\mu\rightarrow e\gamma)=\frac{3\alpha}{8\pi}(|g_L|^2+|g_R|^2)\,,
\eea
where 
\bea
g_{L,R}=\eta_{L,R}\sum_i (U^{ei}_{L,R}U^{i\mu\dagger}_{L,R}) 
F(x^i_{L,R})\,,
\eea
where
\bea
F(x^i_{L,R})=
\left[ \frac{x^i_{L,R}(1-5x^i_{L,R}-2(x^{i}_{L,R})^2)}{2(x^i_{L,R}-1)^3}+
\frac{3(x^{i}_{L,R})^2}{(x^i_{L,R}-1)^4}\ln x^i_{L,R} \right]\,.
\label{g}
\eea
In these expressions $\eta_L=1,$ $\eta_R=M^2_{W_L}/M^2_{W_R},$  
$x^i_{L,R}=(m^i/M_{W_{L,R}})^2$ and summation goes over all neutrino
masses $m^i.$ 

The function $F(x)$  varies from
0 to 1 as $x$ goes from 0 to $\infty.$ Therefore, only the heavy neutrinos 
contribute to the $\mu$ decay and  we can constrain
the  light-heavy and heavy-heavy mixings only.     
Functions $g_L$ and $g_R$ represent  contributions from the 
left-handed and right-handed interactions, respectively. 
In the case of $M_{W_R}=1.4$ TeV and neutrino masses as high as 
argued in  Section 3 one has 
$F(x_L)\approx 1$ and $F(x_R)\approx 0.98$
which implies that there is no significant suppression due to 
the function $F$ in the right-handed sector. However, the
suppression appears due to the parameter $\eta$ which is small
in the case of right-handed currents. On the other hand, since our
model allows $|U_R|^{e\mu}$ to be  of order one,
 we still can bound the heavy-heavy mixings even 
 for quite large $W_R $ masses.
For fixed neutrino mixing angles, 
the branching ratio for the process approaches a constant value when neutrino
masses are sufficiently large. 
If we let $v_R$ to grow then the light-heavy mixings approach zero and the
amplitude of the process vanishes. 
 Due to the mixing  between left- and right-
handed gauge bosons one would also expect to get  terms  proportional  
to neutrino mass but  these terms vanish 
in the case of real photon.

The present experimental bound on the branching ratio for the process is
$B(\mu\rightarrow e\gamma)< 4.9\cdot10^{-11}$ \cite{br1}.
This result is not sensitive to the left- and right-handed contributions
separately. Therefore, in order to constrain mixings in the left and right
sector we assume that only one of them is dominant at the time.
Limits obtained in this way are optimistic ones, if  both 
terms contribute the bounds would be more stringent.
Therefore, the following limits  should be taken as  suggested ones 
 just to get a feeling of what one can  expect in this
class of models.

We assume for simplicity that the 
heavy neutrino masses are similar in magnitude.
Then, in the first approximation,
 the values of $F(x)$ can be taken to be
equal for all neutrino generations  and 
we can easily constrain the quantity
$ |U_{L,R}|^{e\mu}=\left(\sum_i  U^{ei}_{L,R}U^{i\mu\dagger}_{L,R}
\right)^{\frac{1}{2}}.$ Here the summation goes over the heavy neutrinos
only which means that in the case of $U_L$  we can bound light-heavy
mixings and  in the case of $U_R$ heavy-heavy mixings. As expected,
the process is not sensitive to light-light mixings.   
In Fig.2 we plot the constrains on $|U_{L}|^{e\mu}$ and 
in Fig.3  the constrains on $|U_{R}|^{e\mu}$
from the searches of the decay $\mu\rightarrow e \gamma$ 
(curves denoted by $a$) as functions of the heavy neutrino mass.
In Fig.3 we take
$M_{W_R}=1.4$ TeV. As can be seen, the curves become almost constant
for neutrino masses  allowed by the perturbative unitarity.
The limiting value for $ |U_{L}|^{e\mu},$ which does not depend on 
$M_{W_R}$ but on the sensitivity of the experiment only,      
converges  at $|U_{L}|^{e\mu}=0.016$ which agrees with similar
bounds obtained in other models \cite{tommde}. 
The bound on $|U_{R}|^{e\mu} ,$
however, depends almost linearly  on  the right gauge boson mass. 
In order to
get limits on $|U_{R}|^{e\mu}$ for other values of $M_{W_R}$ one 
has to scale the curve $a$ in Fig.3  by an appropriate factor.
With the present value of $M_{W_R}$ the neutrino mass should
exceed 0.6 TeV in order to  bound $|U_{R}|^{e\mu}$ below unity.
If the mass of  $W_R$ exceeds 5 TeV then  the sensitivity of the 
experiment  is not sufficient to constrain $|U_{R}|^{e\mu}$
any more even for the maximally allowed neutrino masses.

In the future  more stringent constaraints on 
$|U_{L,R}|^{e\mu}$  can be obtained from $\mu\rightarrow e\gamma$
searches. According to the proposals
\cite{prop}   new experiments will not only rise the sensitivity
but also use almost 100\% polarized muons. This will allow one to
get direct information about the right- and left-handed interaction
contributions to the decay  and, thus, constrain the model further.  
At present the curves denoted by $a$ in Figs.2,3   imply, indeed, 
  severe constraints on our model.

However, one can do better even with the present experimental data.
The extraordinary sensitivity of the experiments looking for $\mu-e$
conversion in nuclei together with the fact that the branching ratio
of the conversion grows with 
the heavy neutrino mass will provide us
with  constraints that are stronger than those from $\mu\rightarrow e\gamma.$

The $\mu-e$ conversion in nuclei is induced by the flavour changing
$Z\bar{e}\mu$ current which can be parametrized as
\bea
J^{\mu}_{Z\bar{e}\mu}=\frac{g^3}{(4\pi)^2\cos\theta_W}
\bar{e}\gamma^\mu (f_L P_L+f_R P_R) \mu\,,
\label{cur}
\eea  
where $g$ is the weak coupling constant and $P_{R,L}=(1\pm\gamma_5)/2.$
The corrections from photon exchange to $\mu-e$ conversion are small
in the case of heavy neutrinos \cite{conv} since the $\gamma\bar{e}\mu$
amplitude does not grow with the  heavy neutrino mass 
 and we ignore them. We also neglect the contributions from
the new right-handed $Z_R$ exchange and diagrams involving $\Delta^{++}$
 since they are  supressed by the $Z_R$ and $\Delta^{++}$
masses, respectively. 
In this approximation the process receives contributions from
the loop diagrams in Fig.4.  
We have calculated the parameters $f_{L,R}$ of the  effective vertex
(\ref{cur}) providing the results
\bea
f_{L,R}(x^i_{L,R}) &=& \frac{\rho_{L,R}}{2} \left[ \sum_i
(U^{ei}_{L,R}U^{i\mu\dagger}_{L,R}) \left( \frac{(x^i_{L,R})^2-6x^i_{L,R}}
{2(x^i_{L,R}-1)}+\frac{3(x^i_{L,R})^2+2x^i_{L,R}}{2(x^i_{L,R}-1)^2}
\ln x^i_{L,R}\right.\right. \nn \\
& & \left.
-\frac{3x^i_{L,R}}{4(x^i_{L,R}-1)}+ \frac{(x^i_{L,R})^3-2(x^i_{L,R})^2+
4x^i_{L,R}}{4(x^i_{L,R}-1)^2}\ln x^i_{L,R} \right) \nn \\
& &+ \sum_{ij}
(U^{ei}_{L,R}(U^{\dagger}_{L,R}U_{L,R})^\ast_{ij}U^{j\mu\dagger}_{L,R}) 
\left( \frac{x^i_{L,R}x^j_{L,R}}{2(x^i_{L,R}-x^j_{L,R})}\ln\frac{x^j_{L,R}}
{x^i_{L,R}}\right) \nn \\
& & + \sum_{ij}
(U^{ei}_{L,R}(U^{\dagger}_{L,R}U_{L,R})_{ij}U^{j\mu\dagger}_{L,R}) 
\sqrt{x^i_{L,R}x^j_{L,R}}\left( \frac{1}{4} \right. \nn \\
& &-\frac{(x^i_{L,R})^2-4x^i_{L,R}}{4(x^i_{L,R}-1)(x^i_{L,R}-x^j_{L,R})}
\ln x^i_{L,R} \nn \\
& & \left.\left.
-\frac{(x^j_{L,R})^2-4x^j_{L,R}}{4(x^j_{L,R}-1)(x^j_{L,R}-x^i_{L,R})}
\ln x^j_{L,R} \right)\right]\,,
\label{fz}
\eea
where $\rho_{L}=1,$ $\rho_R=\eta/(1-\eta)$ and summation goes over
all neutrino species. We see from these expressions that the leading
terms go as $x^i\ln x^i.$ Therefore the branching ratio of $\mu-e$
conversion in nuclei grows as the fourth power of the heavy
neutrino mass and the contribution from the light neutrinos 
is negligible. 
The right-handed current contribution
is supressed by the factor of $\rho_R$ and also by the smaller values of 
$x_R$ if compared with $x_L.$   Eq.(\ref{fz}) has terms both proportional
to the second and fourth power of the mixing matrices which are multiplied
by $x^i$ in the leading order. 
In the light of our analysis of the process $\mu\rightarrow e\gamma,$
which constrains the mixings well below unity, we conclude that
only the terms containing $|U|^2$ are relevant in our analysis. 
For the left-handed terms this is obvious also from the theoretical point
of view since the light-heavy mixings cannot exceed $k/v_R.$ 
We emphasise that, unlike in the $\gamma\bar{e}\mu$ interaction case,
if $v_R\rightarrow\infty$ then the current (\ref{cur})
 does not vanish and approaches 
a constant value proportional to the electroweak breaking scale.

The best limit on the flavour changing current $J^{\mu}_{Z\bar{e}\mu}$
arises from the search for $\mu-e$ conversion in nuclei.
With the general couplings defined in Eq.(\ref{cur})
the total nuclear muon capture rate for nuclei with 
atomic number $A\lsim 100$ is \cite{conv}
\bea
B\approx\frac{G_F^2\alpha^3}{\pi^2}m^3_\mu p_e E_e\frac{Z^4_{eff}}{Z}
|F(q)|^2\frac{2}{\Gamma_{capture}}(f_L^2+f_R^2)Q_W^2\,,
\label{rat}
\eea
where $p_e$ and $E_e$ are the electron momentum and energy,
$E_e\approx p_e\approx m_\mu$ for the present process, $F(q)$
is the nuclear form factor, measured for example in electron scattering
processes
\cite{fq} and $Z_{eff}$ has been  determined in Ref.\cite{zeff}.
The coherent nuclear charge $Q_W=(2Z+N)v_u+(Z+2N)v_d$ associated with 
the vector current of nucleon is a function of 
the  quark couplings $v_{u,d}$ to the $Z$ boson and 
the nucleon charge and atomic number  $A=Z+N.$
For $^{48}_{22}Ti$ we use the experimental data $\Gamma_{capture}=
(2.590\pm 0.012)\cdot 10^6\;\mbox{{\rm s}}^{-1}$ \cite{br2},
$F(q^2\sim m^2_{\mu})=0.54$ and $Z_{eff}=17.6.$ 
With the present experimental bound $B(\mu-e)=4\cdot 10^{-12}$ 
 \cite{br2} the resulting limit
on the flavour changing couplings $f^2_{L,R}$
is 
\bea
(f_L^2+f_R^2)<2.6\cdot 10^{-13}.
\eea

From this constraint we get bounds on the flavour changing
neutrino mixings. Again, we assume that either the left- or right-handed
currents dominate when bounding the  light-heavy or hevy-hevy mixings.
 Since only the heavy states contribution is
significant we take  for simplicity the masses of heavy neutrinos 
to be almost degenerate i.e. $|m_i^2-m_j^2|/(m_i^2+m_j^2)\ll 1.$
In this case, the bounds on $|U_{L}|^{e\mu}$ and $|U_{R}|^{e\mu}$  
obtained from $\mu-e$ conversion in nuclei are ploted in Fig.2
and Fig.3, respectively, and denoted by $b.$
As previously, we take $M_{W_R}=1.4$ TeV.  
Indeed, these bounds on $|U_{L}|^{e\mu}$ are always 
more stringent than the ones from 
$\mu\rightarrow e\gamma$ 
and become very restrictive (of the order of $10^{-5}$) if the neutrino 
 masses approach the maximum value allowed by perturbative unitarity.
If we rise the mass of  $W_R$  then also $m_i$ can be bigger
and the bound on $|U_{L}|^{e\mu}$ approaches zero for $m_i\rightarrow
\infty.$ 
Let us note that the indirect limit  $|U_{L}|^{e\mu}< 0.063$ 
is worser than those in Fig.2.

As can be seen in Fig.3, even with the present lower bound on $M_{W_R}$
we have a lower bound of 1.1 TeV on the heavy neutrino mass in
order to bound   $|U_{R}|^{e\mu}$ below unity.
For $m_N<1.3$ TeV the best limit on $|U_{R}|^{e\mu}$ comes from the decay 
$\mu\rightarrow e\gamma.$ For heavier neutrino the $\mu-e$ conversion
limit becomes more restrictive approaching a few times $10^{-3}$
in the case of maximally allowed neutrino masses.   
The right-handed current contribution to the branching ratio of 
$\mu-e$ conversion is suppressed by the factor $\rho_R$ 
but, due to the almost linear dependence on $m_N,$ 
this suppression is partly compensated by the large values of neutrino 
mass. Therefore, quite large values of $M_{W_R}$ for which we still can
constrain   $|U_{R}|^{e\mu}$ for some neutrino masses
are allowed. In this case we
cannot bound the heavy-heavy neutrino mixings if $M_{W_R}$ exceeds 200 TeV.

For  completness we have also analysed the process
$\mu\rightarrow ee^+e^-.$ In the left-right model it arises from 
  tree level diagram of $\Delta^{++}$ exchange and loop
diagrams which involve heavy neutrinos and gauge bosons.
The contribution from  the tree level graph has been analysed in 
Ref.\cite{meee}. Assuming that $\Delta^{++}$ is very heavy and 
the dominant contribution comes from the loop diagrams we have
calculated the branching ratio of the process. In addition to
the $J^{\mu}_{Z\bar{e}\mu}$ current sub-diagrams, there are 
also box diagrams 
which give rise
to the quadratic dependence with the neutrino mass.
The bounds on off-diagonal neutrino mixings derived from the
experimental limit $B(\mu\rightarrow ee^+e^-)<1.0\cdot 10^{-12}$
\cite{eeee} are complementary to the bounds from
$\mu\rightarrow e\gamma$ and $\mu-e$ conversion in nuclei but
the limits are about 2-3 times less stringent than those from the $\mu-e$
 conversion. Therefore, we do not present our lengthy 
expressions here. We note that this result agrees with similar
ones in Refs.\cite{tommde,conv}.

\section{Conclusions}

We have shown that the mixings between light and heavy
neutrinos in $SU(2)_L\times SU(2)_R\times U(1)_{B-L}$ models 
with the discrete left-right symmetry (\ref{trans}) can be naturally
significant. Minimizing the full scalar potential of the 
theory and avoiding  fine tunings of the parameters of the 
potential, we obtain the relation $v_L v_R\simeq k_1k_2$ for the vevs.
This leads naturally to a  singular neutrino mass matrix allowing
to explain the smallness of the ordinary neutrino masses and keep
the mixings with heavy states large at the same time. 
We emphasise that, unlike in the SM extended with heavy right-handed neutrinos
where one has to fine tune the neutrino Yukawa couplings,
the singularity of the mass matrix in the left-right models
 may come even if all the Yukawa
couplings are equal. 
We have derived constraints on the intergenerational neutrino mixings
from the searches for the rare flavour changing processes. 
These processes are  sensitive not only to the mixings of light and
heavy neutrinos but, due to the right-handed charged currents 
present in the model, also to  the off-diagonal mixings between
heavy neutrinos.    

For the   allowed values of neutrino masses the best constraints
on light-heavy mixings occur from $\mu-e$ conversion in nuclei. 
If the masses of heavy neutrinos are below $\sim 1.3$ TeV and
$M_{W_R}=1.4$ TeV 
then the best constraints on
 heavy-heavy mixings come from the decay  
$\mu\rightarrow e\gamma.$  For heavier neutrinos, up to
the perturbative unitarity bound, the best limits
arise from the data of $\mu-e$ conversion in nuclei. 
The effective $Ze\mu$ coupling, which induces the conversion
process, shows a strong non-decoupling behaviour being quadratically 
dependent  of the heavy neutrino mass   while the amplitude of the process  
$\mu\rightarrow e\gamma$ is almost constant for the neutrino masses 
above the electroweak scale. Therefore, the heavier the neutrinos,
the more stringent are the constraints from $\mu-e$ conversion.
 In our model, the perturbative unitarity requirement relates 
the upper bound on the mass of heavy neutrino to the mass of 
new right-handed gauge boson as given in Eq.(\ref{mn2}). 
Consequently, rising the right-handed scale 
the constraints on light-heavy mixings can be
 more  stringent  than e.g. in the SM with singlet neutrinos 
where the heavy neutrino masses are limited to be below 1 TeV.

Since the right-handed charged currents  are suppresses
by the large mass of $W_R$ also
the constraints on off-diagonal heavy-heavy mixings 
depend on $M_{W_R}.$ The present experiments on $\mu\rightarrow e\gamma$ 
can constrain  the heavy-heavy mixings only if $M_{W_R}< 5$ TeV.
In the case of $\mu-e$ conversion, however, this suppression is much 
weaker  and the corresponding bound is 200 TeV.

The bounds obtained from the analyses of decay $\mu\rightarrow ee^+e^-$
are complementary to the previous ones but about two times less
stringent than those from the $\mu-e$ conversion in nuclei. 
This result agrees with the results in  
 Refs.\cite{tommde,conv}.

The searches for flavour changing processes  
imply, indeed, very strong constraints on our model. 
The planned experiments looking for $\mu-e$ conversion
and $\mu\rightarrow e\gamma$ with polarized muons are specially 
suitable for finding signals arising from the type of models
we have considered.

\subsection*{Acknowledgement}

We thank J. Bernab\'eu,  A. Santamar\'{\i}a and D. Tommasini 
for clarifying discussions.
G.B. acknowledges the Spanish Ministry of
Foreign Affairs for a MUTIS fellowship and M.R. thanks the
Spanish Ministry of Science and Education for a postdoctoral
grant at the University of Valencia. This work is supported by CICYT under 
grant AEN-93-0234.

\newpage

\section*{Figure captions}

\begin{itemize}

\item[{\bf Fig.1.}]
Feynman diagrams which involve  heavy neutrinos and contribute
to  the decay $\mu\rightarrow e\gamma$. \\

\item[{\bf Fig.2.}]
Constraints on the off-diagonal mixing $|U|_{e\mu}$  between 
 light and heavy neutrinos as 
 functions of the heavy neutrino mass derived from $\mu\rightarrow e\gamma$ 
and $\mu-e$ conversion in nuclei and denoted by $a$ and $b$, respectively.
\\

\item[{\bf Fig.3.}]
Constraints on the intergenerational mixing $|U|_{e\mu}$
among   heavy neutrinos as 
functions of the heavy neutrino mass derived from $\mu\rightarrow e\gamma$ 
and $\mu-e$ conversion in nuclei and denoted by $a$ and $b$, respectively.
The mass of the right-handed gauge boson is taken to be 1.4 TeV.
\\

\item[{\bf Fig.4.}]
Feynman diagrams which involve heavy neutrinos and contribute to  
the $\mu-e$ conversion in nuclei.
\\

\end{itemize}

\newpage

\begin{figure*}[hbtp]
\begin{center}
 \mbox{\epsfxsize=12cm\epsfysize=8cm\epsffile{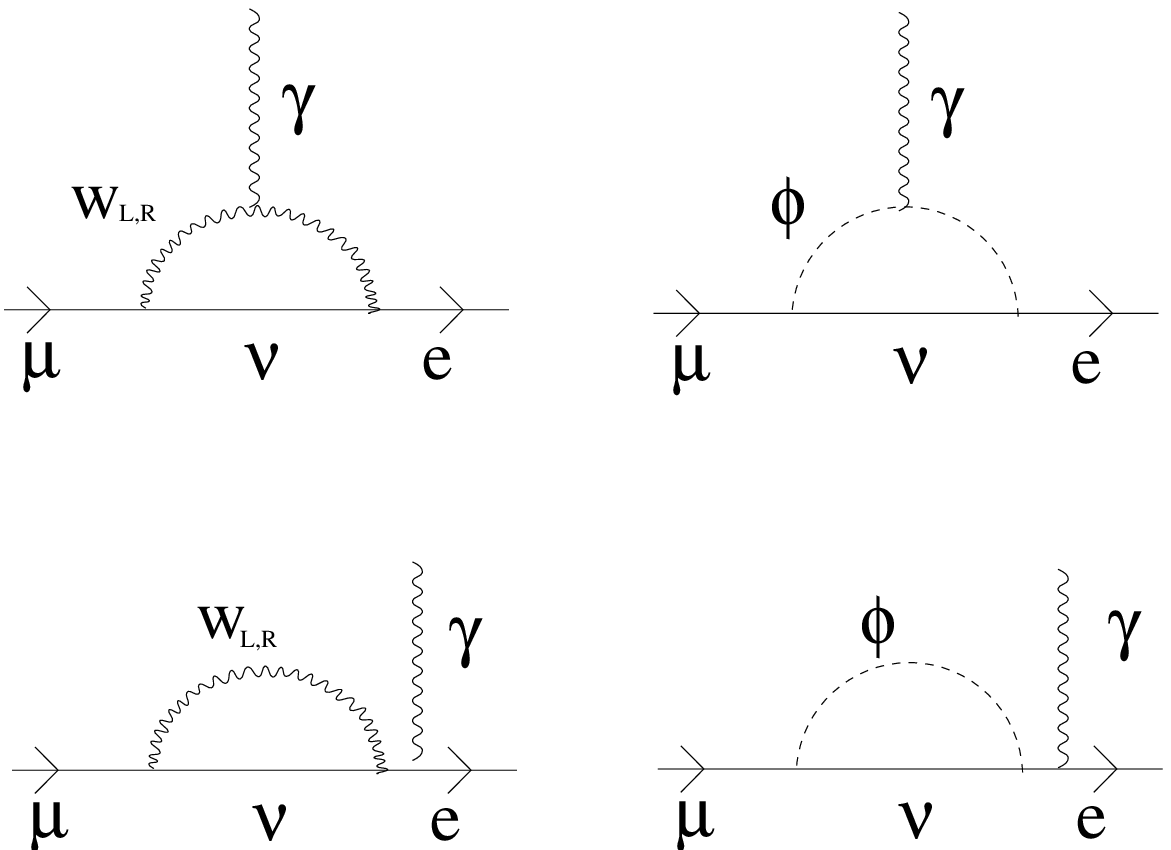}}
\caption{
}
\end{center}
\end{figure*}

\begin{figure*}[hbtp]
\begin{center}
 \mbox{\epsfxsize=12cm\epsfysize=12cm\epsffile{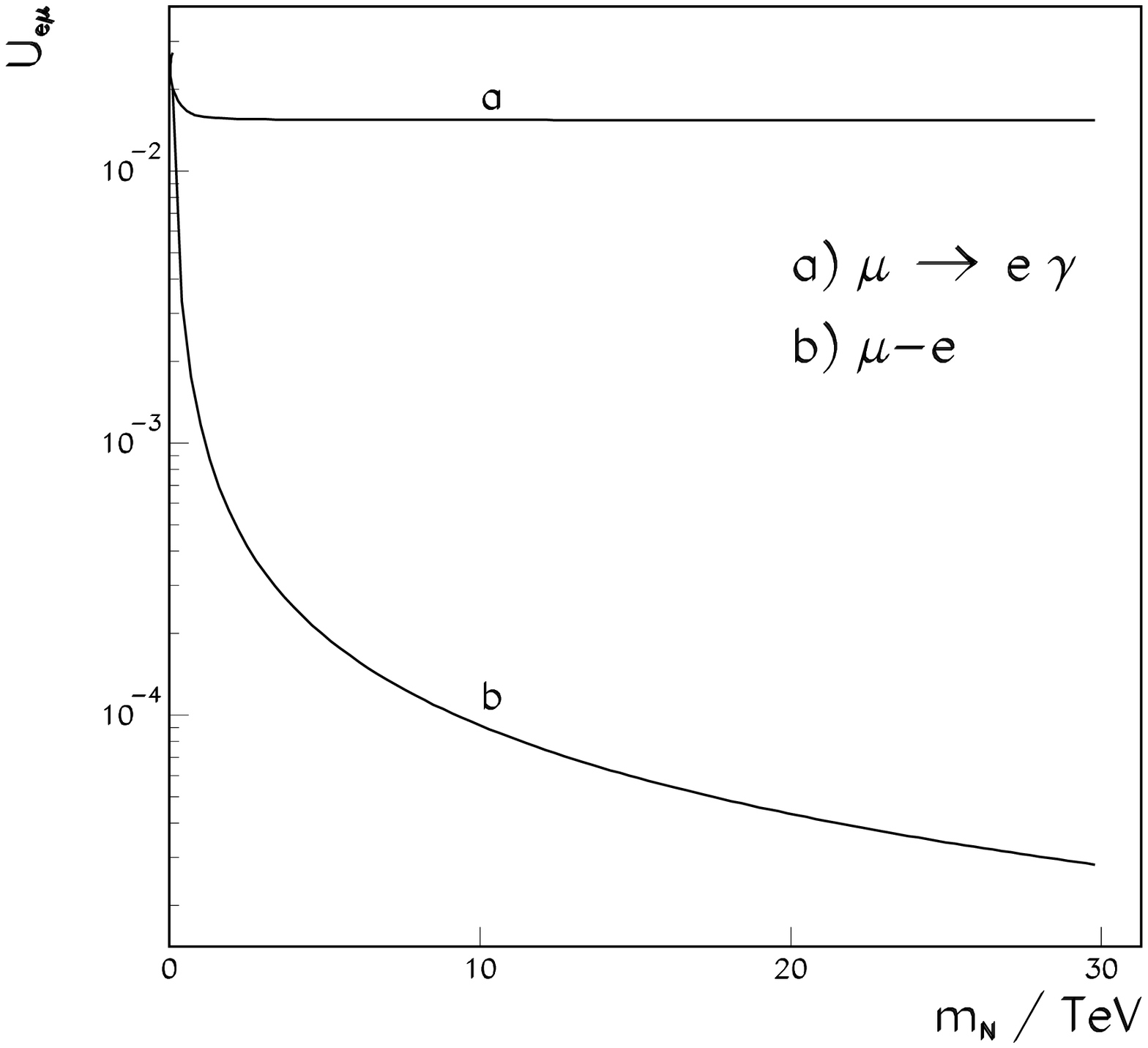}}
\caption{
}
\end{center}
\end{figure*}

\begin{figure*}[hbtp]
\begin{center}
 \mbox{\epsfxsize=12cm\epsfysize=12cm\epsffile{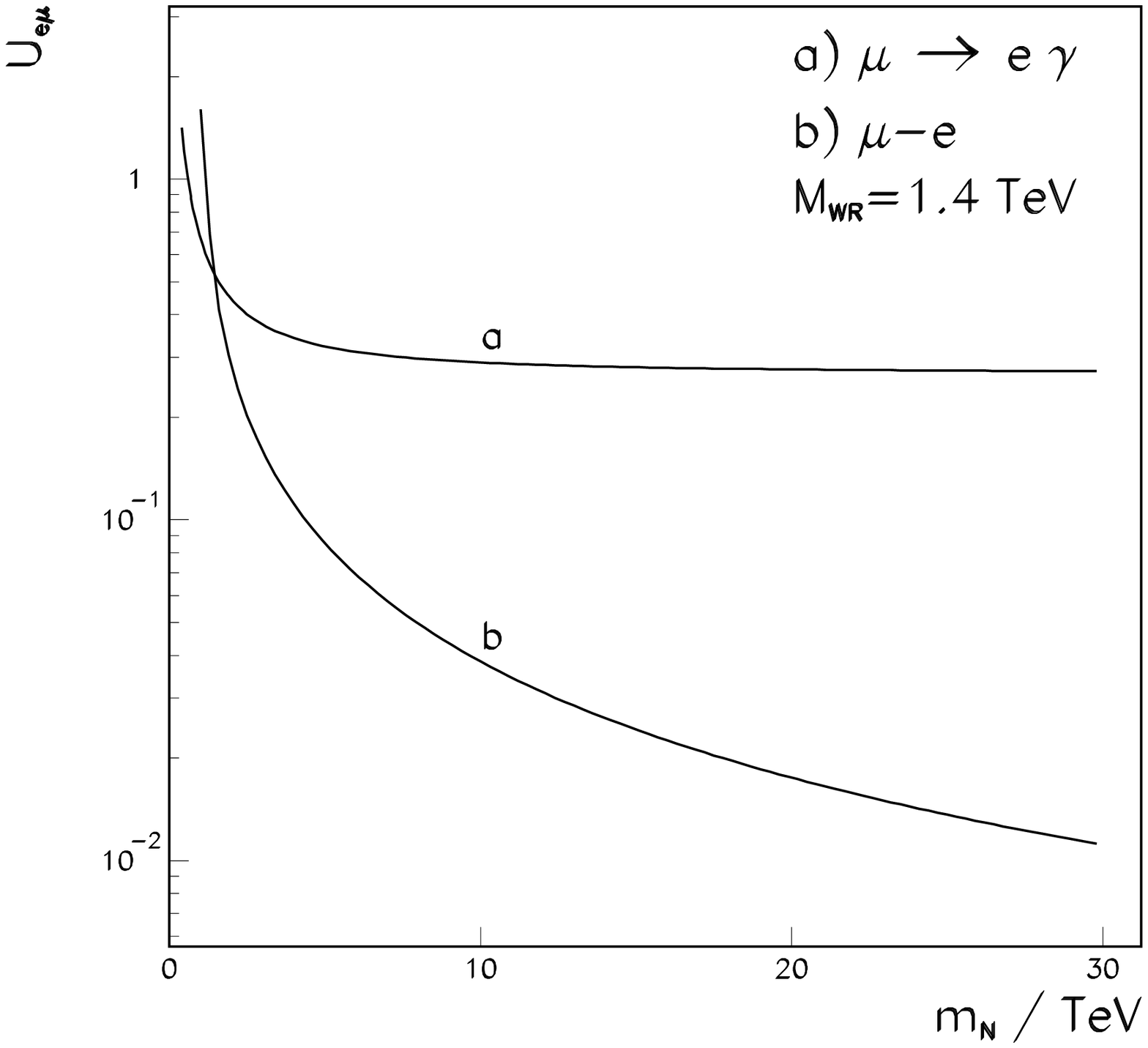}}
\caption{
}
\end{center}
\end{figure*}

\begin{figure*}[hbtp]
\begin{center}
 \mbox{\epsfxsize=12cm\epsfysize=8cm\epsffile{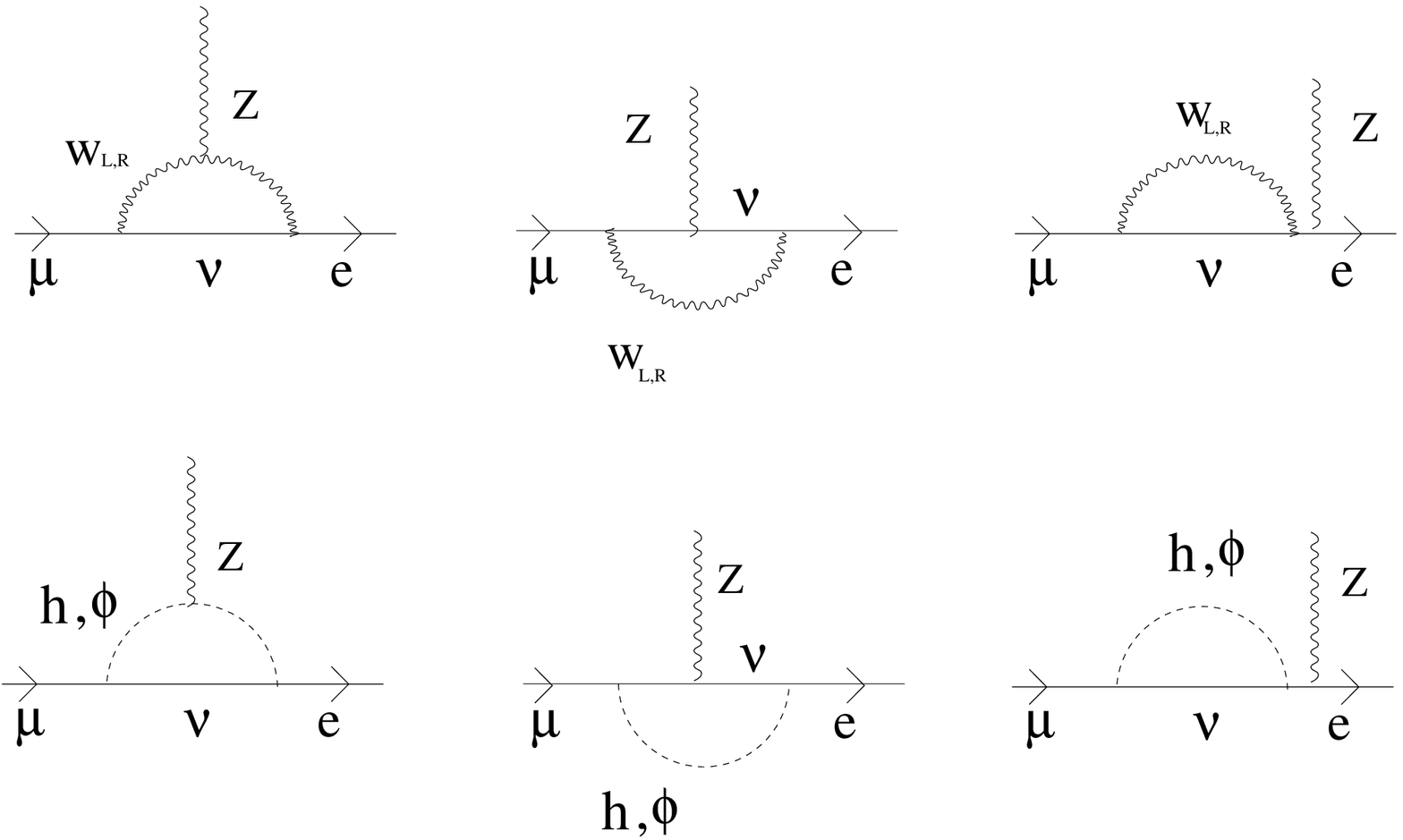}}
\caption{
}
\end{center}
\end{figure*}


\begin{thebibliography}{99}

\bibitem{sol}
K.S. Hirata et al., \prl{66} (1990) 1301; \\
A.I. Abazov et al., \prl{67} (1991) 3332; \\
K. Nakamura, Nucl. Phys.  (Proc. Suppl.) {\bf B31} (1993).

\bibitem{atm}
K.S. Hirata et al., \pl{B280} (1992) 146; \\
D. Casper et al., \prl{66} (1993) 2561.

\bibitem{cobe}
G.F. Smoot et al., Ast. J. {\bf 396} (1992) L1.

\bibitem{lang}
P. Langacker and D. London, \pr{D38} (1988) 886; \\
E. Nardi and E. Roulet, \pl{B248} (1990) 139; \\
G. Bhattacharyya et al., \prl{64} (1990) 2870; \\
C.P. Burgess et al., \pr{D49} (1994) 6115; \\
G. Bhattacharyya, \pl{B331} (1994) 143.

\bibitem{precision}
E. Nardi, E. Roulet and D. Tommasini, \pl{B327} (1994) 316
and  \\ \pl{B344} (1995) 225.

\bibitem{lang2}
P. Langacker and S.U. Sankar, \pr{D40} (1989) 1569; \\
G. Barenboim, J. Bernabeu and M. Raidal, FTUV/96-51, submitted 
to \pr{D}.

\bibitem{senj2} G. Senjanovic and A. Sokorac, \np{B164} (1980) 305.

\bibitem{nondec}
T. Appelquist and J. Carazzone, \pr{D11} (1975)  2856.

\bibitem{loop}
M. Veltman, \np{B123} (1977) 89;\\
J. Collins, F. Wilczek and A. Zee, \pr{D18} (1978) 242; \\
C.S. Lim and T. Inami, \ptp{65} (1981) 297; \\
C.S. Lim and T. Inami, \ptp{67} (1982) 69; \\
J.D. Vergados, \prep{133} (1986) 1; \\
T.P. Cheng and L.F. Li, \pr{D44} (1991) 1502; \\ 
D. Tommasini et al., \np{B444} (1995) 451.

\bibitem{see-saw}
T. Yanagida, \ptp{B135} (1978) 66; \\
M. Gell-Mann, P. Ramond and A. Slansky, in {\em Supergravity},
eds. P. van Nieuwenhuizen and D. Freedman (North-Holland, 1979),
p. 315. 


\bibitem{wyler}
D. Wyler and L. Wolfenstein, \np{B218} (1983) 205.


\bibitem{esingl}
R.N. Mohapatra and J.F.W. Valle, \pr{D34} (1986) 1642; \\
J. Bernabeu et al., \pl{B187} (1987) 303; \\
J.L. Hewett and T. Rizzo, \prep{183} (1989) 193; \\
E. Nardi, \pr{D48} (1993) 3277.


\bibitem{buch}
W. Buchm\"uller and C. Greub, \np{B363} (1991) 345.


\bibitem{lr} J. C. Pati and A. Salam, Phys. Rev. {\bf D10} (1975) 275; \\
R. N. Mohapatra and J. C. Pati, Phys. Rev. {\bf D11} (1975) 566 and 2558; \\
G. Senjanovic and R. N. Mohapatra, \pr{D12} (1975) 1502.

\bibitem{ss} R. N. Mohapatra and G. Senjanovic, Phys. Rev. Lett. {\bf 44} 
(1980) 912 and \\
 Phys. Rev. {\bf D 23} (1981) 165.

\bibitem{tommde}
See D. Tommasini et al., in Ref.\cite{loop}.

\bibitem{wmass} G. Beall, M. Bander and A. Soni, Phys. Rev. Lett. {\bf 48}
(1982) 8484.

\bibitem{gun} J. F. Gunion, J. Grifols, A. Mendez, B. Kayser and F. Olness,
Phys. Rev. {\bf D40} (1989) 1546; \\
N. G. Deshpande, J. F. Gunion, B. Kayser and F. Olness,
Phys. Rev. {\bf D44} (1991) 837; \\
K. Huitu, J. Maalampi, A. Pietil\"a and M. Raidal, HU-SEFT R 1996-16,
eprint hep-ph/9606311.

\bibitem{gabriela} 
D. Chang, \np{B214} (1983) 435;\\
G. Ecker and W. Grimus, \np{B258} (1985) 328;\\
J.M. Frere et al., \pr{D46} (1992) 337; \\
G. Barenboim and J. Bernab\'eu, preprint  FTUV/96-9,
e-print hep-ph/9603379, \\ to appear in \zp{C}; \\
G. Barenboim, J. Bernab\'eu and M. Raidal, preprint  FTUV/96-24, 
eprint hep-ph/9608445, \\ to appear in \np{B}.

\bibitem{pdb} {\it Review of Particle Properties}, \pr{D54} (1996) 1.

\bibitem{senj} 
G. Senjanovic, \np{B153} (1979) 334.

\bibitem{lalo}
P. Langacker and D. London, \pr{D38} (1988) 907.

\bibitem{t22}
M.C. Gonzalez-Garcia and J.F.W. Valle, \mpl{A7} (1992) 477; \\
A. Ilakovac and A. Pilaftsis, \np{B437} (1995) 491.

\bibitem{moha}
R.N. Mohapatra, \pr{D46} (1992) 2990.

\bibitem{br1}
LAMPF collaboration, R.D. Bolton et al., \prl{56} (1986) 2461.

\bibitem{prop}
Y. Kuno and Y. Okada, \prl{77} (1996) 434.

\bibitem{conv}
J. Bernab\'eu, E. Nardi and D. Tommasini, \np{B409} (1993) 69.

\bibitem{fq}
B. Frois and C.N. Papanicolas, Ann. Rev. Nocl. Sci. {\bf 37} (1987) 133.

\bibitem{zeff}
H.C. Chiang et al., \np{A559} (1993) 526.

\bibitem{br2}
T. Suzuki, D.F. Measday and J.P. Roalsvig, \pr{C35} (1987) 2212.

\bibitem{meee}
M.L. Swartz, \pr{D40} (1989) 1521; \\
M. Lusignoli and S. Petrarca, \pl{B226} (1989) 397.

\bibitem{eeee}
SINDRUM collaboration, U. Bellgardt et al., \np{B299} (1988) 1.


\end{thebibliography}
\end{document}